\documentclass[aps,twocolumn,prl]{revtex4-1}
     
\usepackage{graphicx}
\usepackage{dcolumn}
\usepackage{bm}
\usepackage{txfonts}
\usepackage{color}
\usepackage{amssymb}

\begin{document}
\preprint{APS/123-QED}

\title{Multiple metamagnetic quantum criticality in Sr$_3$Ru$_2$O$_7$}

\author{Y. Tokiwa$^{1,2}$}
\author{M. Machalwat$^1$}
\author{R. S. Perry$^3$}
\author{P. Gegenwart$^{1,4}$}

\affiliation{$^1$I. Physikalisches Institut, Georg-August-Universit\"{a}t G\"{o}ttingen, 37077 G\"{o}ttingen, Germany}
\affiliation{$^2$Department of Physics, Kyoto University, Kyoto 606-8502, Japan}
\affiliation{$^3$London Centre for Nanotechnology, Faculty of Maths \& Physical Sciences, University College London, London - WC1E 6BT, UK}
\affiliation{$^4$Experimentalphysik VI, Center for Electronic Correlations and Magnetism, Augsburg University, 86159 Augsburg, Germany}

\date{\today}

\begin{abstract}
Bilayer strontium ruthenate Sr$_3$Ru$_2$O$_7$ displays pronounced non-Fermi liquid behavior at magnetic fields around 8~T, applied perpendicular to the ruthenate planes, which previously has been associated with an itinerant metamagnetic quantum critical end point (QCEP).  We focus on the magnetic Gr\"uneisen parameter $\Gamma_{\rm H}$, which is the most direct probe to characterize field-induced quantum criticality. We confirm quantum critical scaling due to a putative two-dimensional QCEP near 7.845(5)~T, which is masked by two ordered phases A and B, identified previously by neutron scattering. \textcolor{black}{In addition we find evidence for a QCEP at 7.53(2)~T and determine the quantum critical regimes of both instabilities and the effect of their superposition.}
\end{abstract}

\maketitle
Quantum criticality denotes critical behavior that is associated with continuous transformations of matter at zero temperature. Due to the absence of thermal fluctuations at $T=0$ it is qualitatively different from classical criticality~\cite{sachdev:qcp-book}. In metals the unconventional excitation spectrum near a quantum critical point (QCP) causes the breakdown of Fermi liquid (FL) behavior and its intimate relation to exotic states, such as unconventional superconductivity, adds even more importance to this topic. To date, the influence of quantum critical magnetic excitations on electrons in a metal is far from being understood. For instance the applicability of the itinerant Hertz-Millis-Moriya theory on $f$-electron based Kondo lattice systems has been disproved by several experiments~\cite{gegenwart-review} and alternative descriptions are not fully established yet. Quantum criticality related to itinerant metamagnetism is exceptional in the sense, that electronic degrees of freedom are irrelevant, and a quantitative application to experimental results should be possible~\cite{millis:prl-02}.

The generic metamagnetic quantum critical end point (QCEP) arises from the suppression to $T=0$ of the end point of a line of first-order metamagnetic transitions in temperature-field phase space by tuning e.g. composition, pressure or the magnetic field orientation~\cite{millis:prl-02}. Metamagnetic QCEPs have been realized in the $f$-electron based compounds CeRu$_2$Si$_2$~\cite{Daou-prl06,Weickert-prb10} and UCoAl~\cite{Aoki-JPSJ11}, as well as $d$-electron Sr$_3$Ru$_2$O$_7$~\cite{Grigera-prb03,Rost-Science09}.

We focus on bilayer strontium ruthenate Sr$_3$Ru$_2$O$_7$. Magnetization of this compound along the tetragonal $c$-axis at low temperature exhibits three successive super-linear, i.e. metamagnetic, rises at $\mu_0H_{\rm M1}$=7.5\,T, $\mu_0H_{\rm M2}$=7.85\,T and $\mu_0H_{\rm M3}$=8.1\,T~\cite{Perry-prl04}. The first one is a metamagnetic cross-over (M1). The second and third ones are first order metamagnetic transitions (M2 and M3), ending at critical temperatures of about 1 and 0.5 K, respectively~\cite{Grigera-Science04}. A line of second-order thermal phase transitions, connecting the critical end points of M2 and M3, has been discovered in electrical resistivity and thermodynamic experiments~\cite{Grigera-Science04,gegenwart-prl06}, which recently by neutron scattering has been identified as phase boundary of a spin-density-wave (SDW) "phase A"~\cite{Lester-nmat15,Borzi-science07} (see Fig.~1.). The lower and upper critical fields of SDW-A correspond respectively to $H_{\rm M2}$ and $H_{\rm M3}$. Additionally, another SDW "phase B" has been observed in between $H_{\rm M3}$ and 8.3\,T~\cite{Lester-nmat15,Stingl-pss13}. \textcolor{black}{The observed incommensurate ordering vectors in both SDW phases have been related to Fermi surface nesting~\cite{Lester-nmat15}.} Magnetic susceptibility and magnetostriction have revealed the strongest peak at the M2 metamagnetic transition and weaker maxima at M1 and M3~\cite{Grigera-Science04}. \textcolor{black}{The critical field has been extrapolated to $\mu_0H_{c2}=7.845(5)$~T~\cite{gegenwart-prl06}, which is indeed very close to $\mu_0H_{\rm M2}$}. Non Fermi liquid (NFL) behavior at elevated temperatures was previously associated to a critical field close to $H_{\rm M2}$~\cite{Perry-prl04}. Outside the SDW phases A and B and not to close to the M1 crossover, thermal expansion obeys quantum critical scaling in accordance with the expectations for a two-dimensional (2D) metamagnetic QCEP near 7.845~T~\cite{gegenwart-prl06}. This includes both the predicted divergence upon cooling within the quantum critical regime as well as the magnetic field dependence within the low-temperature FL regime upon tuning the field from both sides towards M2. \textcolor{black}{However, the previous description of the specific heat coefficient $C/T$ by a strong divergence $|H_{\rm M2}-H|^{-1}$~\cite{Rost-Science09,Rost-PSS10} is in clear contradiction to the theoretical prediction $C/T \sim |H_c-H|^{-1/3}$~\cite{millis:prl-02}.}

\textcolor{black}{We solve this discrepancy by proving that Sr$_3$Ru$_2$O$_7$ displays two QCEPs at $\mu_0H_{c1}=7.53(2)$~T and $\mu_0H_{c2}=7.845(5)$~T, respectively. We determine regimes in phase space where either of the two QCEPs leads to scaling of the magnetic Gr\"uneisen parameter. We also show where scaling fails due to the superposition of criticality from both instabilities. Multiple quantum criticality as origin for behavior that is incompatible with the generic predictions of QCPs can be of relevance for various material classes.}

\begin{figure}
\includegraphics[width=\linewidth,keepaspectratio]{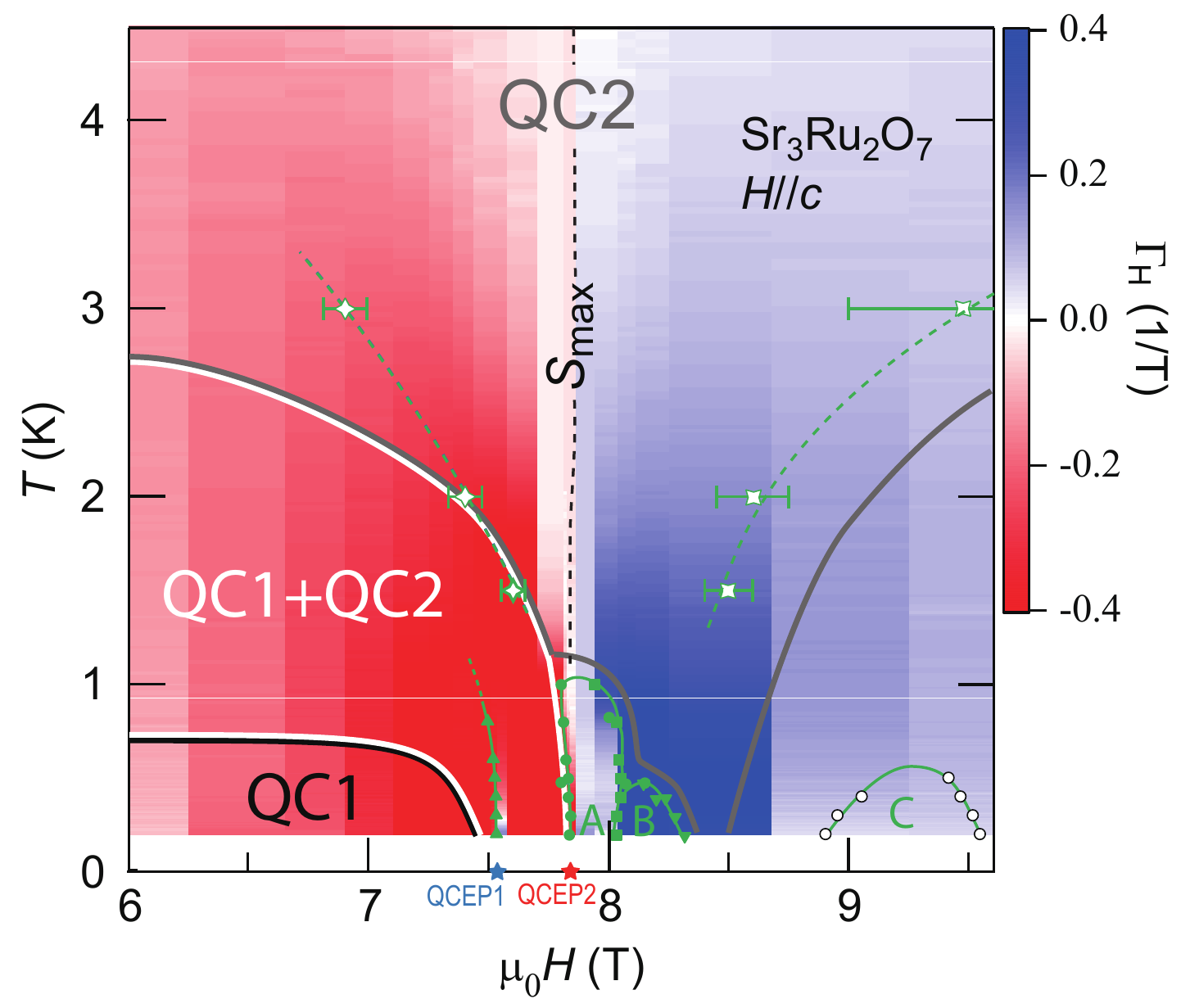}
\caption{Phase diagram of Sr$_3$Ru$_2$O$_7$ for $H\parallel c$ with color-coding of the magnetic Gr\"{u}neisen parameter, $\Gamma_{\rm H}$. Solid green symbols mark positions of sharp peaks in $\Gamma_{\rm H}(H)$, related to metamagnetism~\cite{Grigera-Science04} and the spin-density-wave phases A and B~\cite{Lester-nmat15}. Open green symbols indicate positions of maxima in the field dependence of specific heat. The dotted black line marks $\Gamma_{\rm H}$=0,  corresponding to a local entropy maximum. The stars on the x-axis show the positions of the two metamagnetic quantum critical end points \textcolor{black}{QCEP1 and QCEP2. Grey, white and black solid lines bound different regimes. Labels "QC1" and "QC2" denote regions where quantum critical scaling with respect to QCEP1 and QCEP2 is observed. Within the "QC1+QC2" regime scaling fails due to the superposition of criticality from both instabilities, see supplemental material (SM) ~\cite{SM}). Anomalies in isothermal $\Gamma_{\rm H}(H)$ scans are indicated as yet unidentified regime "C".}}
\end{figure}

The magnetic Gr\"{u}neisen parameter, $\Gamma_{\rm H}=T^{-1}(dT/dH)_S$ measures the relative temperature change with magnetic field under adiabatic conditions, called adiabatic magnetocaloric effect. Due to the entropy accumulation near field-driven quantum criticality, generically this property is expected to obey (i) a sign change when tuning the field across the critical value, (ii) a divergence upon cooling (at constant field) within the quantum critical regime~\cite{zhu,garst-prb05} and (iii) universal scaling within both the FL and quantum critical regime. The adiabatic magnetocaloric effect can be accurately determined with the aid of the alternating-field method~\cite{tokiwa-rsi11}. Using this technique several field-induced quantum critical points have been characterized~\cite{tokiwa-prl09,tokiwa-prl13,tokiwa-prl13_2}. The magnetic Gr\"uneisen parameter provides direct access to the critical exponents which characterize quantum criticality. Below, we report a thorough study of $\Gamma_{\rm H}$, determined by the alternating field technique~\cite{tokiwa-rsi11}, as well as heat capacity measurements performed with the quasi-adiabatic heat pulse technique, on a high-quality single crystal of Sr$_3$Ru$_2$O$_7$, grown by the floating zone technique~\cite{Perry-JCG04}, for fields applied along the $c$-axis.

\begin{figure}
\includegraphics[width=\linewidth,keepaspectratio]{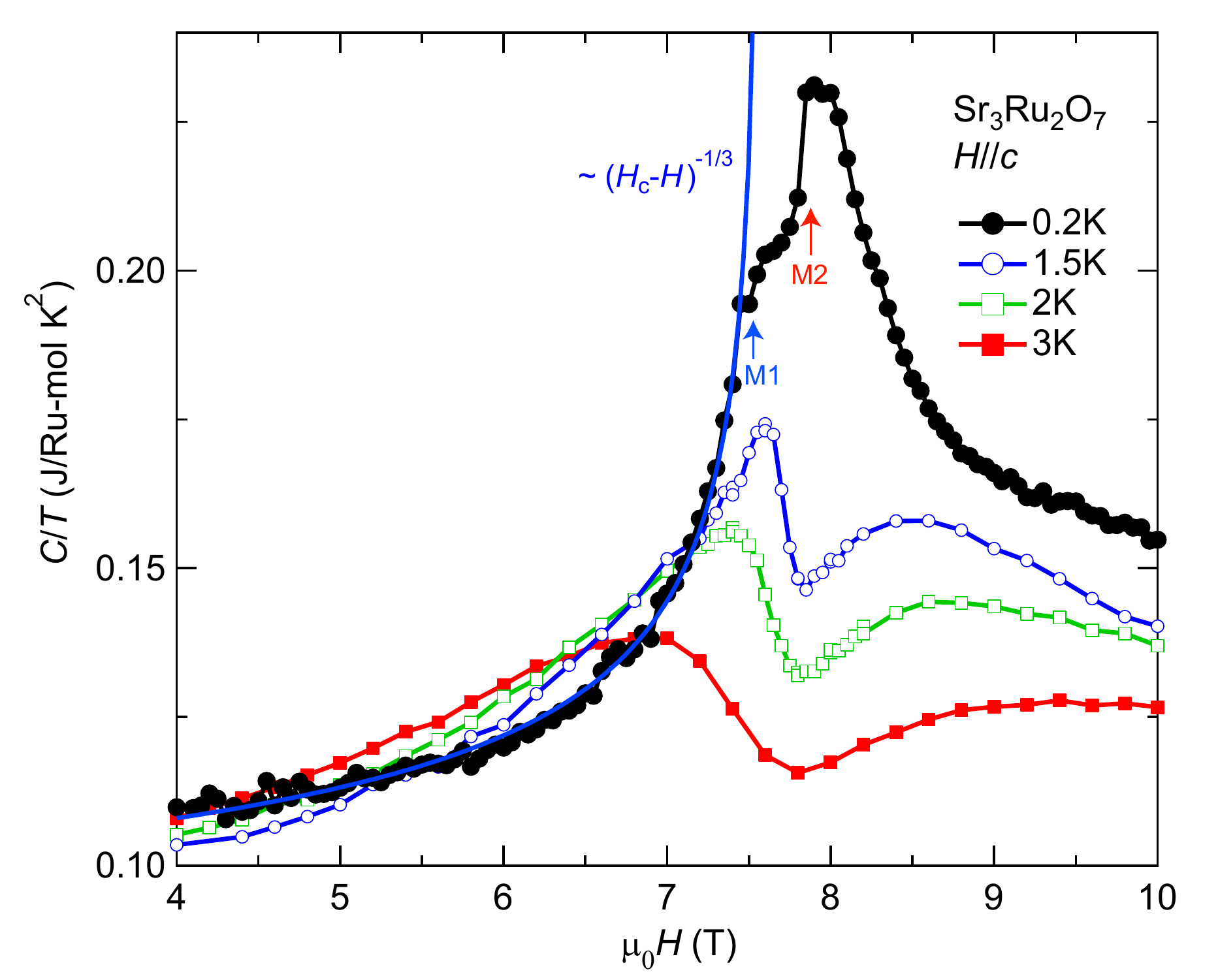}
\caption{(Color online) Specific heat divided by temperature of Sr$_3$Ru$_2$O$_7$ as a function of magnetic field applied parallel to $c$-axis at different constant temperatures. Solid blue line indicates $C/T=\alpha/[\mu_0(H_c-H)]^{1/3}+\gamma$ with $\alpha=(0.073\pm0.002)$\,J/Ru-mol$\cdot$ K$^2$T$^{1/3}$, $\mu_0H_c=7.57(4)$\,T and $\gamma=0.058(1)$\,J/Ru-mol$\cdot$K$^2$, in accordance with a two-dimensional metamagnetic QCEP~\cite{millis:prl-02,Zacharias-prb13}.}
\end{figure}

Figure 2 displays the magnetic field dependence of the specific heat coefficient at various low temperatures. Data at 0.2~K display a single peak at 7.85~T. At larger temperatures, this peak is split into two peaks and the respective separation increases with increasing temperature. Qualitatively, such behaviour is characteristic to itinerant metamagnetism and has also been found for CeRu$_2$Si$_2$~\cite{Aoki-jpsj01}. For a generic QCEP with a critical free energy $F_{\rm cr}(h)=F_{\rm cr}(-h)$ (where $h=\mu_0(H-H_c)$), symmetric peaks for the heat capacity are expected. Our measurements, however, display more broadened $C/T$ peaks on the high-field compared to the low-field sides. As discussed later, this may be related to a slight increase of the effective dimensionality of the critical fluctuations at large fields.

The magnetic field dependence of the 0.2~K data is in perfect agreement with \textcolor{black}{previous data~\cite{Rost-Science09,Rost-PSS10}, see SM~\cite{SM}. 
As shown by the blue solid line in Fig. 2, the data are well described by $C/T\propto (H_c-H)^{-1/3}$, predicted for a 2D QCEP~\cite{millis:prl-02,Zacharias-prb13} with critical field close to $H_{\rm M1}$ but significantly smaller than $H_{\rm M2}$. This indicates that the previously anticipated scenario with a {\it single} field-tuned QCEP near $H_{\rm M2}$~\cite{Rost-Science09} is insufficient.}

The existence of two separate 2D metamagnetic QCEPs is evident from the analysis of the magnetic Gr\"{u}neisen parameter $\Gamma_{\rm H}$ given below. In contrast to the specific heat coefficient, which \textcolor{black}{has a substantial non-critical background, $\Gamma_{\rm H}$ is more sensitive to quantum criticality because of a negligibly small non-critical contribution.}

\begin{figure}
\includegraphics[width=\linewidth,keepaspectratio]{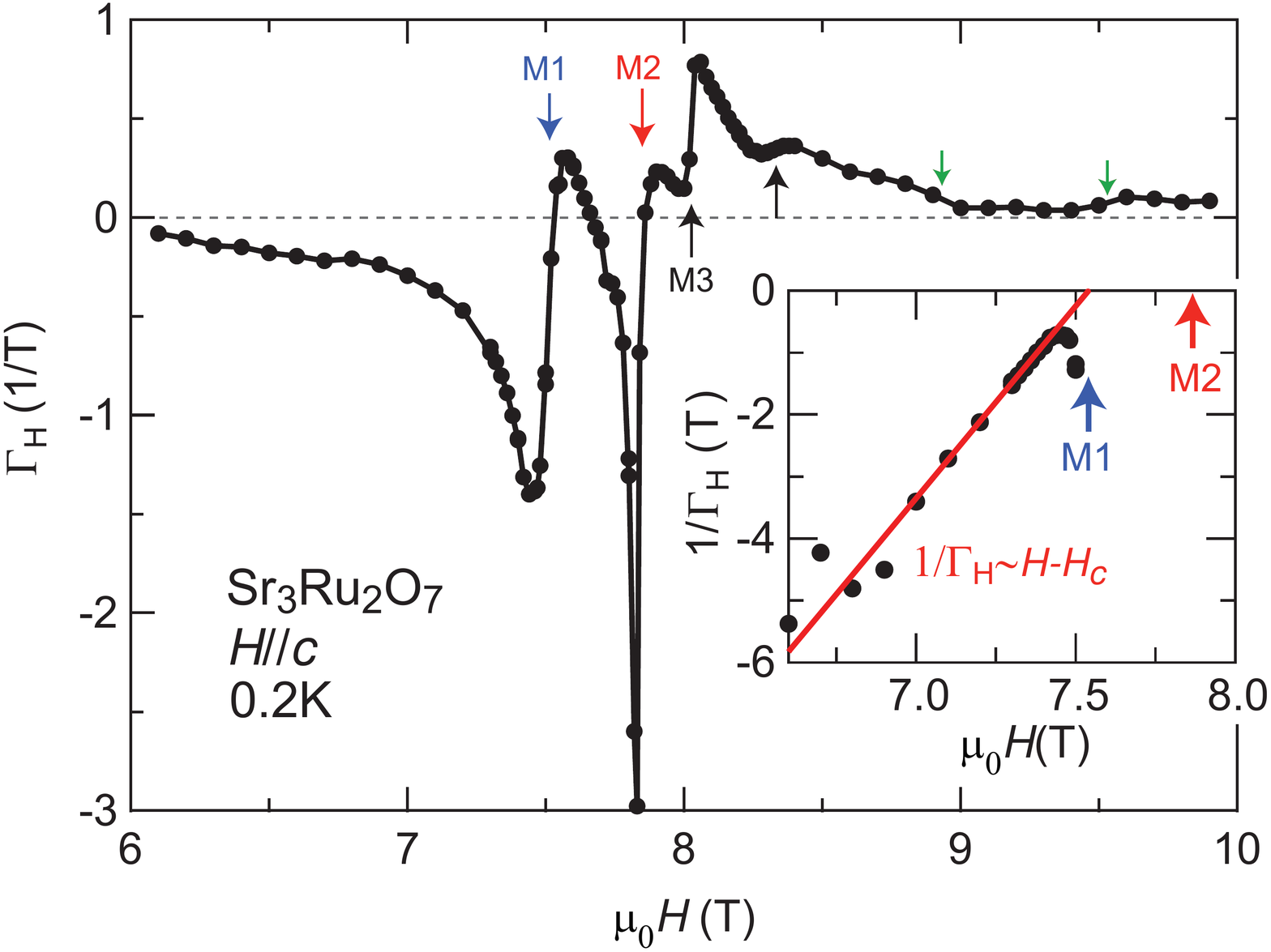}
\caption{(Color online) Magnetic field dependence of the magnetic Gr\"{u}neisen parameter $\Gamma_{\rm H}$ at 0.2\,K of Sr$_3$Ru$_2$O$_7$. The field is applied parallel to the $c$-axis. Arrows at $\mu_0H_{\rm M1}$=7.5\,T, $\mu_0H_{\rm M2}$=7.85 and $\mu_0H_{\rm M3}$=8.0 indicate metamagnetic anomalies. Black arrow at 8.3\,T marks the anomaly related to SDW-B phase~\cite{Stingl-pss13,Lester-nmat15}. Green arrows indicate anomalies which correspond to open circles in Fig.~1, enclosing an anomalous regime "C". Inset shows a plot of $1/\Gamma_{\rm H}$ vs $\mu_0H$. Solid red line represents a linear fit, $1/\Gamma_{\rm H}=-\mu_0(H-H_c)/G_r$ with $G_r=-0.17(1)$ and $\mu_0H_c=7.51(2)$\,T.}
\end{figure}

Figure~3 shows an isothermal scan of the magnetic Gr\"{u}neisen parameter at 0.2~K. $\Gamma_{\rm H}(H)$ increases by more than a factor 10 in between 6 to 7.5~T. For any field-tuned QCP, the magnetic Gr\"{u}neisen parameter displays a generic $1/h$ divergence~\cite{zhu}. Thus, the inverse of the Gr\"{u}neisen parameter versus field must follow a linear dependence and crosses zero at the critical field. As shown in the inset of Fig. 3, this universal dependence is indeed observed, yielding a critical field very close to $H_{\rm M1}$, which confirms our heat capacity analysis.

At fields beyond $H_{\rm M1}$ a cascade of further sign changes and anomalies is found in $\Gamma_{\rm H}(H)$. They are associated with metamagnetic transitions M2 and M3 and respcetively the SDW phases A and B~\cite{Stingl-pss13,Lester-nmat15}, as well as (see the green arrows) an anomaly labeled "C" in the phase diagram of Fig. 1, whose magnetic Gr\"{u}neisen parameter signature is discussed in SM~\cite{SM}.

Each zero-crossing of $\Gamma_{\rm H}(H)$ from negative to positive with increasing field indicates an entropy accumulation which arises either above a QCP or at the boundary of an ordered phase. Although the behavior is very complex, it is qualitatively similar to the field dependence of the low-temperature thermal expansion coefficient~\cite{gegenwart-prl06}. A simpler field dependence with only one sign change of $\Gamma_{\rm H}(H)$ related to M2, is found at elevated temperatures above 1~K\textcolor{black}{~\cite{SM}. There, the thermodynamic properties are mostly influenced by QCEP2 (cf. Fig.~1).}


\begin{figure}
\includegraphics[width=\linewidth,keepaspectratio]{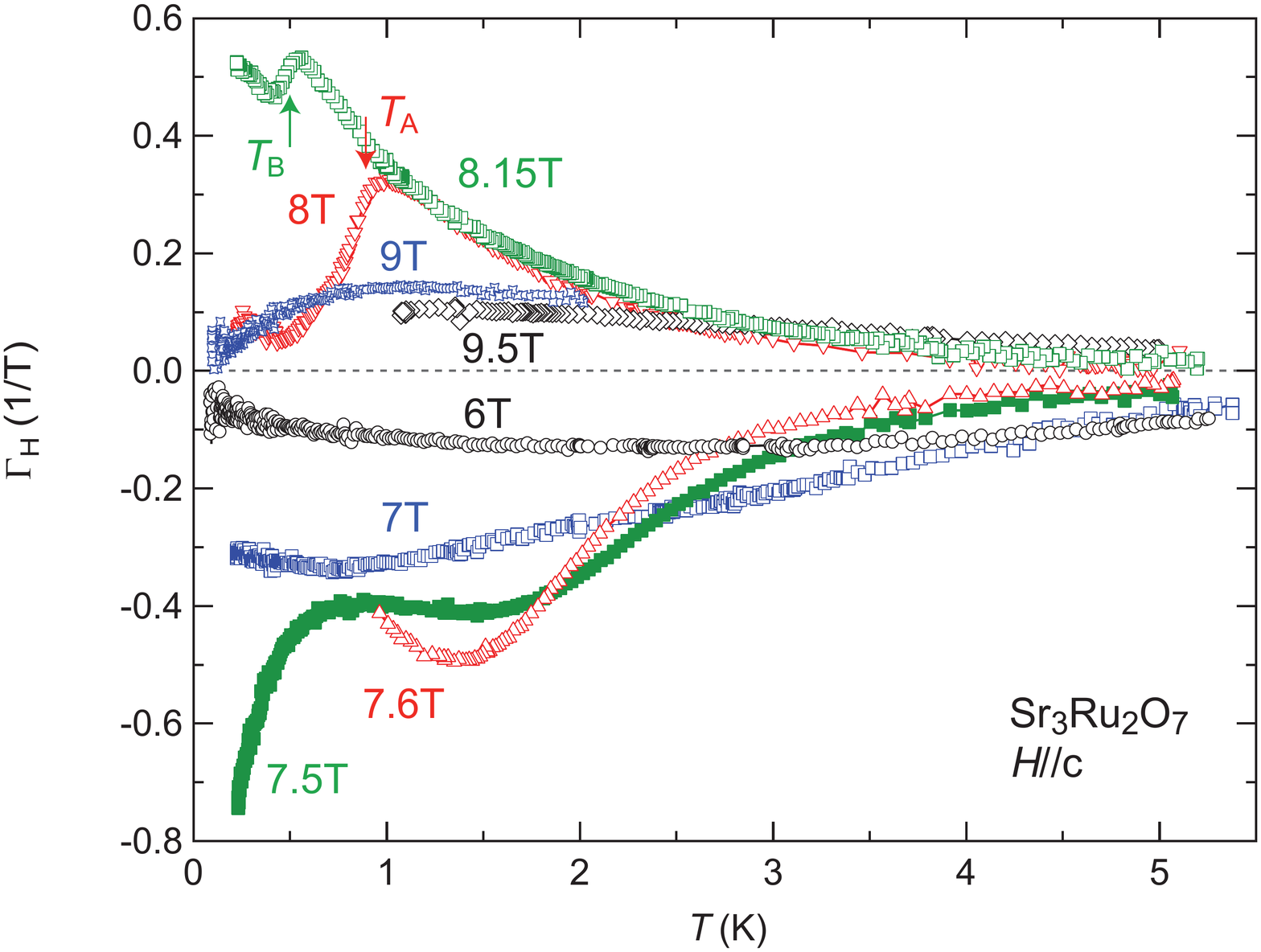}
\caption{(Color online) Magnetic Gr\"{u}neisen parameter $\Gamma_{\rm H}$ of Sr$_3$Ru$_2$O$_7$ as a function of temperature at different magnetic fields, applied parallel to the $c$-axis. The red and green arrows indicate the transitions to the spin-density-wave phases A and B~\cite{Lester-nmat15}.}
\end{figure}

In addition to isothermal measurements, we also study the temperature dependence of $\Gamma_{\rm H}$ at various fields, cf. Figure~4. At $T>1$~K, all curves below $H_{\rm M2}$ show a negative $\Gamma_{\rm H}$, while it is positive for $H>H_{\rm M2}$. Since $\Gamma_{\rm H}=-(dM/dT)/C$, where the heat capacity $C>0$, this reflects the change of sign in the temperature dependence of the magnetization associated with metamagnetism (ordinary paramagnetic behavior below $H_{\rm M2}$ and field polarized behavior above $H_{\rm M2}$). The overall symmetric behavior of $\Gamma_{\rm H}(T)$ with respect to the critical field of QCEP2, is reflecting the Ising symmetry of critical metamagnetic fluctuations~\cite{Zacharias-prb13}. Upon cooling, $|\Gamma_{\rm H}|$ increases within the critical regime of QCEP2 and passes a maximum upon entering the low-temperature FL state, as seen e.g. for the 9~T data in Fig. 4. Transitions to phases A and B lead to distinct anomalies indicated by arrows. Particularly interesting behavior is found at 7.5~T where upon cooling $\Gamma_{\rm H}(T)$ passes the minimum at 1.5~K, due to the FL crossover of QCEP2, but subsequently displays a negative divergence as $T\rightarrow 0$, related to the nearby QCEP1 \textcolor{black}{(cf. Fig.~1).}

\begin{figure}
\includegraphics[width=\linewidth,keepaspectratio]{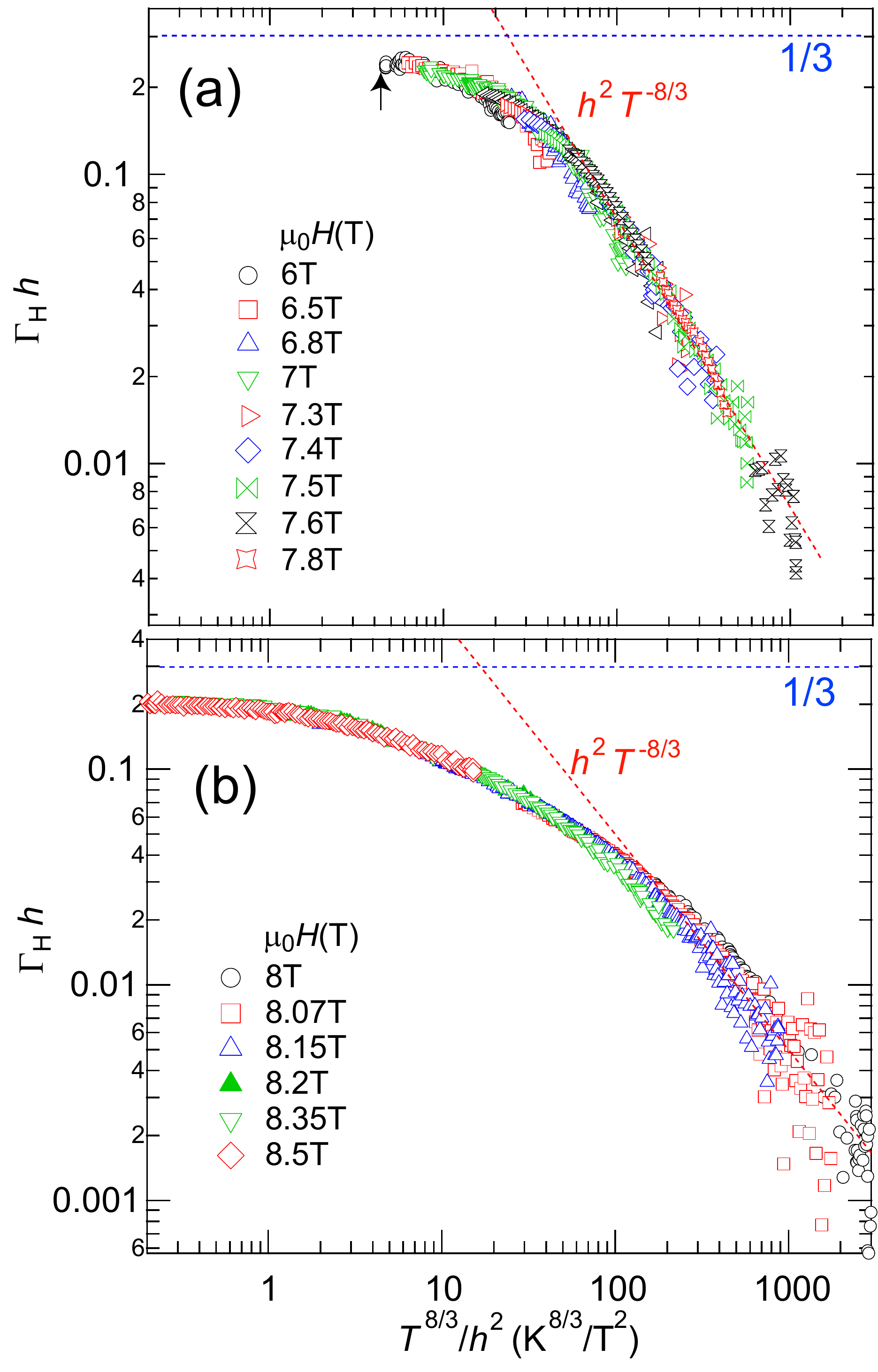}
\caption{(Color online) Metamagnetic quantum critical scaling of the magnetic Gr\"{u}neisen parameter in Sr$_3$Ru$_2$O$_7$. The y-axis displays $\Gamma_{\rm H}h$ while the x-axis shows $T^{8/3}/h^2$, with $h=\mu_0(H-H_{c2})$ and the critical field $H_{c2}=7.845$~T~\cite{gegenwart-prl06}. Panels (a) and (b) displays regimes below and above the critical field. Red and blue dotted lines indicate predicted asymptotic quantum critical and FL behavior for a $d=2$ metamagnetic QCEP~\cite{Zacharias-prb13}. In panel (a) data on the left of the black arrow have been excluded \textcolor{black}{(for failed scaling see SM~\cite{SM}).}}
\end{figure}

We now turn to a quantitative comparison of our data with the theory of metamagnetic quantum criticality~\cite{millis:prl-02,Zacharias-prb13}. The latter predicts $\Gamma_{\rm H}h\sim h^2/T^{(4+2d)/3}$ in the quantum critical and $\Gamma_{\rm H}h=(3-d)/3$ in FL regime, where $d$ denotes the dimensionality and $h=\mu_0(H-H_c)$. This leads to universal scaling in a plot of $\Gamma_{\rm H}h$ vs $h^2/T^{\epsilon}$, where $\epsilon =(4+2d)/3=8/3$ for $d=2$. Respective scaling behavior of our data is shown in Figure 5. Here we fixed the critical field to 7.845~T~\cite{gegenwart-prl06}, which is the position of QCEP2. The data collapse over several orders of magnitude, similar as previously found for thermal expansion~\cite{gegenwart-prl06}, proves quantum critical behavior and indicates the applicability of the itinerant theory. However, a close inspection provides further information\textcolor{black}{~\cite{SM}}. First, for fields below $H_{\rm M2}$, scaling is cut-off near the crossover to the FL regime. This could be associated to the influence of QCEP1, as discussed above. Second, for fields $H>H_{\rm M2}$ the data within the FL regime approach a saturation of $\Gamma_{\rm H}h\approx 0.2$, which is smaller than the value 1/3 predicted for a QCEP with dimensionality $d=2$~\cite{Zacharias-prb13} and may indicate that the effective dimensionality slightly increases at large fields. The value of 0.2 would correspond to $d_{\rm eff}=2.4$. \textcolor{black}{Metamagnetism in Sr$_3$Ru$_2$O$_7$ is supposed to arise from van Hove singularities near the Fermi level~\cite{Tamai-PRL08}. A change of the de Haas-van Alphen frequencies near 8~T has been ascribed to magnetic breakdown~\cite{Mercure-PRB10}. This could explain the increase of the effective dimensionality of critical fluctuations, deduced from our scaling analysis.}


\textcolor{black}{The different regimes where the magnetic Gr\"uneisen parameter displays scaling with respect to QCEP 1 and QCEP2 are indicated in Fig.~1. In between both regimes neither scaling works, because criticality from both instabilities is adding up (see SM~\cite{SM}).
Next,} we discuss the influence of the ordered phases A and B. In the approach of these phase transitions, $\Gamma_{\rm H}$ data deviate from the expected quantum critical scaling. This could be naturally explained by additional contributions to the free energy arising from classical critical behavior. Furthermore, there is an anomalous depression of $\Gamma_{\rm H}$ at 9~T below 1 K (cf. Fig. 4), which could not be accounted for by the scaling due to QCEP2. The magnetic field dependence of $\Gamma_{\rm H}(T)$ (Fig. 3, see also SM~\cite{SM}) indicates low-temperature anomalies in this field regime, labeled "C" in the phase diagram (Fig. 1). Since heat capacity does not show an anomaly these are rather weak thermodynamic signatures for phase formation. The fields where these anomalies are observed are temperature dependent. Thus, it is unlikely, that these anomalies originate from low frequency quantum oscillations~\cite{Rost-Science09}.

Our measurements of the magnetic Gr\"{u}neisen parameter and specific heat coefficient establish the existence of two itinerant metamagnetic QCEPs in  bilayer strontium ruthenate Sr$_3$Ru$_2$O$_7$ for magnetic fields applied parallel to the $c$-direction. \textcolor{black}{QCEP1 appears at a metamagnetic crossover near 7.5~T while QCEP2, which has already previously been established, is located at 7.845~T. The phase diagram shown in Fig. 1 indicates the scaling regimes "QC1" and "QC2" determined from the magnetic Gr\"uneisen parameter behavior (see also SM~\cite{SM}). While "QC2" is largely extended at elevated temperatures, "QC1" is confined to a narrow regimes close to QCEP1. In between these scaling regimes, there exist a range in phase space, in which scaling fails due to the superposition of criticality from both instabilities. The phase diagram is even richer and contains also two SDW phases A and B~\cite{Lester-nmat15} and some anomalous yet unidentified regime labeled "C". Likely, the observed complexity is related to the complicated electronic structure of this material~\cite{Tamai-PRL08}. The Fermi surface contains several pockets that could give rise to nesting and sheets near a van Hove singularity.} From a general perspective, multiple quantum criticality may be of origin of anomalous behaviors in different material classes, including heavy-fermions and high-$T_c$ superconductors. The Gr\"{u}neisen parameter is ideally suited to disentangle multiple quantum criticality.

 \textcolor{black}{Stimulating discussions with M. Brando, M. Garst and C. Stingl are gratefully acknowledged.}

\bibliography{hf}

\vspace{1 cm}
\textbf{SUPPLEMENTAL MATERIAL}

\section{Field dependence of the specific heat}

\begin{figure}[hb*]
\includegraphics[width=\linewidth,keepaspectratio]{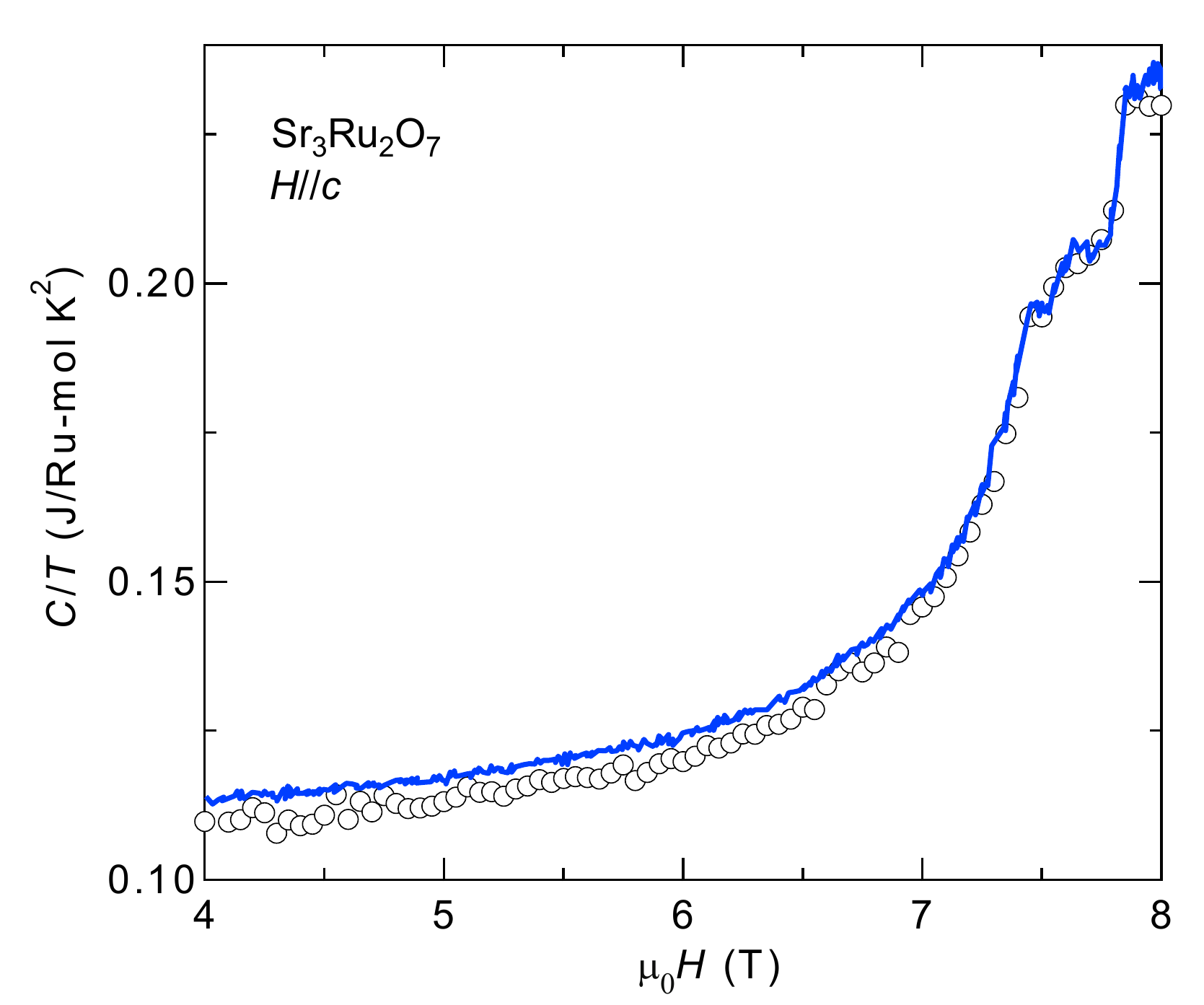}
\caption{Magnetic field dependence of specific heat divided by temperature of Sr$_3$Ru$_2$O$_7$. Open black circles display the data (see main text) measured at 0.2~K, while the solid blue line has been extracted from A.W. Rost et al., Phys. Stat Sol. B {\bf 247}, 513 (2010). The latter data were obtained using an a.c. heat capacity technique during continuous magnetic field sweep at 0.25~K.}
\label{gamma_singlefit}
\end{figure}

In the main text, we are discussing the field dependence of our heat capacity data below 7.5~T. It shows a power-law divergence towards $H_{c1}$ with exponent 1/3 in accordance with the itinerant theory for a two-dimensional QCEP. However, previous data of the field dependence of the entropy increment, obtained by non-adiabatic magnetocaloric effect measurements  in A.W. Rost et al., Science {\bf 325}, 1360 (2009), as well as continuous field sweep a.c. heat capacity data at 0.25~K, see A.W. Rost et al., Phys. Stat Sol. B {\bf 247}, 513 (2010), had been described differently (see main text). Rost et al. have used the function $(H_{c2}-H)^{-1}$ (note the different exponent and different critical field) that would be highly incompatible with the prediction from the itinerant theory. It is therefore interesting to directly compare their data with ours. As shown in Figure~\ref{gamma_singlefit} both data sets for $C/T$ (note that in the Fermi liquid regime $C/T$ is temperature independent) differ by less than 4~mJ/Ru-mol K$^2$ equivalent to 2\%. This nicely indicates the reproducibility of the thermodynamic results on Sr$_3$Ru$_2$O$_7$. Based on our magnetic Gr\"uneisen analysis (cf. inset of Fig. 3 main text), the description of $C/T$ by $(H_{c1}-H)^{-1/3}$ in contrast to $(H_{c2}-H)^{-1}$ is appropriate.

\section{Field dependence of the magnetic Gr\"uneisen ratio}

\begin{figure}[ht*]
\includegraphics[width=\linewidth,keepaspectratio]{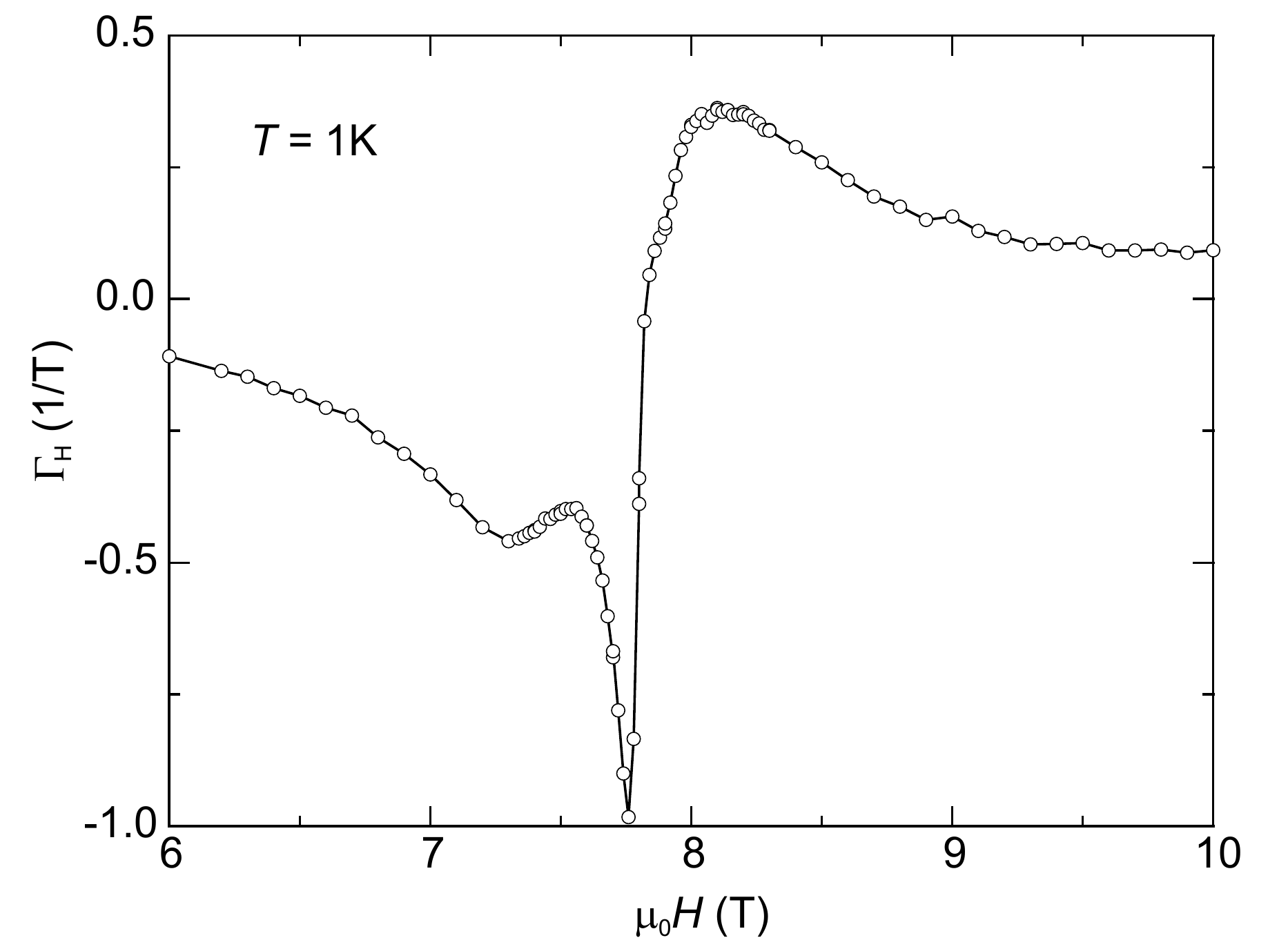}
\caption{Magnetic field dependence of magnetic Grüneisen parameter $\Gamma_H$ of Sr$_3$Ru$_2$O$_7$ at $T=1$\,K.}
\label{1K}
\end{figure}

As shown in Fig.~3 in the main text, at 0.2~K the magnetic Gr\"uneisen ratio displays very complicated behavior with several extrema and sign changes. Figure~\ref{1K} displays respective data taken at 1~K. Upon increasing magnetic field from 6~T, the Gr\"uneisen parameter is negative and increases in absolute value. Near $H_{c1}$ it passes a local minimum and maximum and continues to diverge until a sharp minimum and sign change is reached very close to $H_{c2}$, beyond which $\Gamma_H$ is positive and decreases with increasing $H$. Since $\Gamma_H=-(dS/dH)/C$ the zero crossing of the magnetic Gr\"uneisen ratio, which occurs along a line above the QCEP2 (shown in Fig. 1 of the main text), indicates an accumulation of entropy. This is a generic signature of quantum criticality (cf. M. Garst and A. Roch, Phys. Rev. B {\bf 72}, 205129 (2005)). At lower temperatures, the interplay of QCEP1 and QCEP2 results in a more complicated behavior of $\Gamma_H(H)$ displayed in Fig. 3 of the main text.

\begin{figure}[ht*]
\includegraphics[width=\linewidth,keepaspectratio]{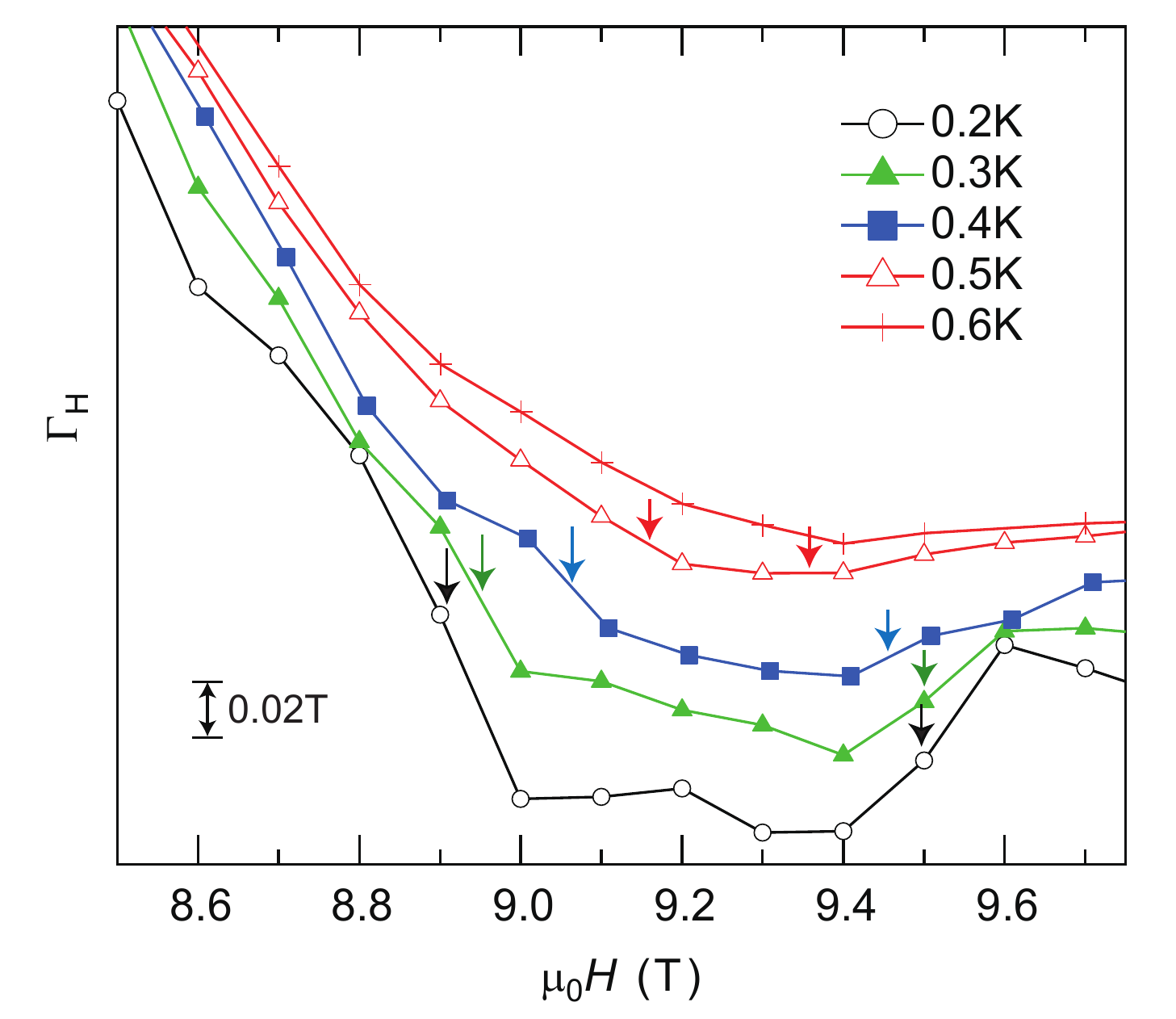}
\caption{Magnetic field dependence of the magnetic Gr\"uneisen parameter $\Gamma_H$ of Sr$_3$Ru$_2$O$_7$ for fields between 8.5 and 9.7~T at different temperatures. Data are shifted vertically for clarity. Arrows indicate inflection points in $\Gamma_H(H)$.}
\label{anomaly}
\end{figure}

Next we focus on isothermal field data of $\Gamma_H(H)$ for large fields between 8.5 and 10~T shown in Figure~\ref{anomaly}. The colored arrows 
indicate anomalous behavior, leading to inflection points in the field dependence. The respective fields are temperature dependent, cf. regime "C" in the phase diagram Fig. 1 of the main text. This excludes quantum oscillations as origin. We note, that no clear signature in heat capacity has been found in this regime of phase space.

\section{Scaling analysis for QCEP1}

\begin{figure}[hb*]
\includegraphics[width=\linewidth,keepaspectratio]{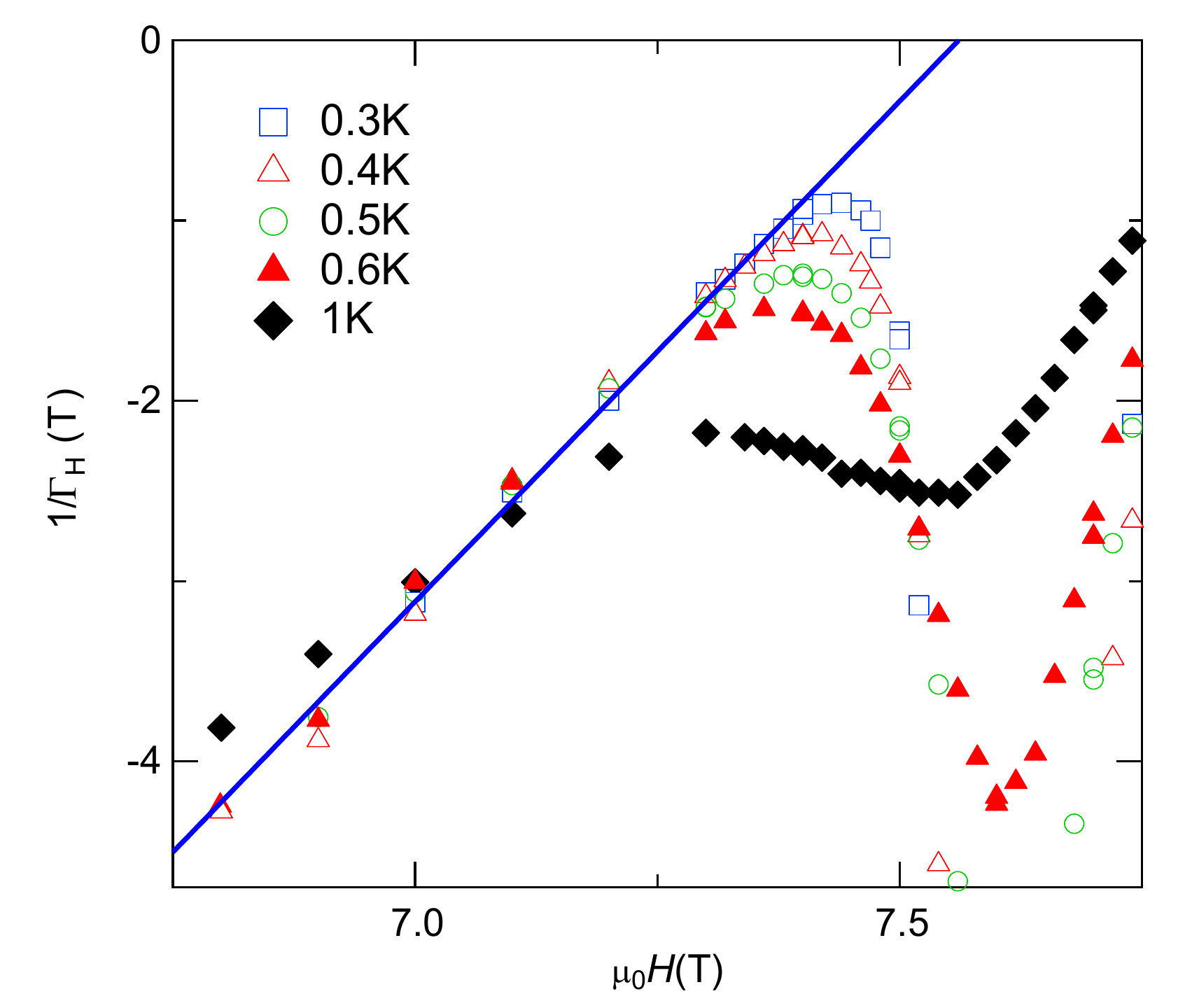}
\caption{Magnetic field dependence of the inverse magnetic Gr\"uneisen parameter for Sr$_3$Ru$_2$O$_7$ at various different temperatures (similar as for 0.2~K plotted in the inset of Fig. 3 main text). The blue line indicates scaling behavior with respect to QCEP1.}
\label{1_G_H}
\end{figure}

\begin{figure}[ht*]
\includegraphics[width=\linewidth,keepaspectratio]{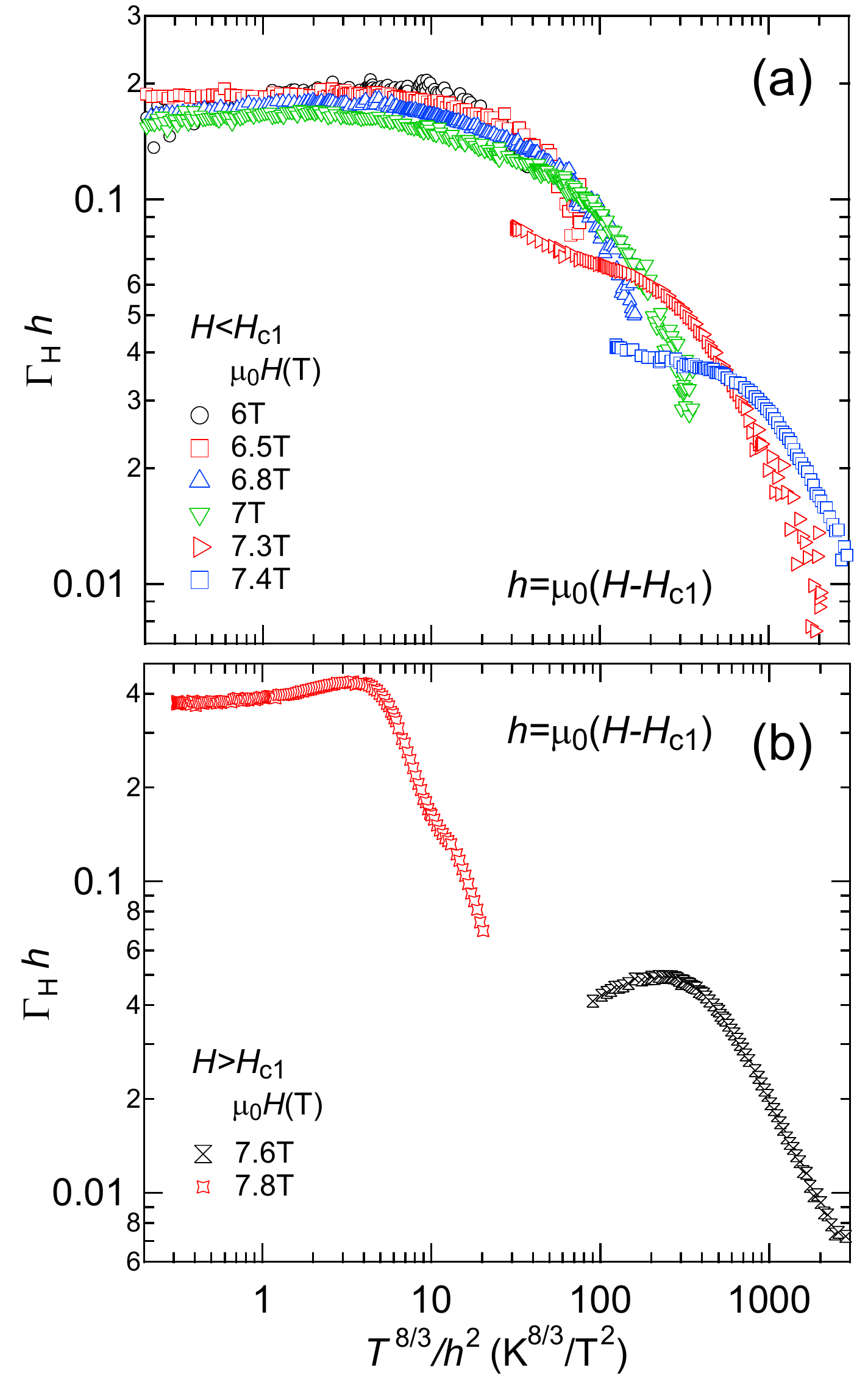}
\caption{Scaling analysis of $\Gamma_H$ for Sr$_3$Ru$_2$O$_7$ with respect to the QCEP1, using $h=\mu_0(H-H_{c1})$ for fields below (a) and above (b) the critical field $\mu_0H_{c1}=7.53(2)$~T.}
\label{scalingHc1}
\end{figure}

\begin{figure}[ht*]
\includegraphics[width=\linewidth,keepaspectratio]{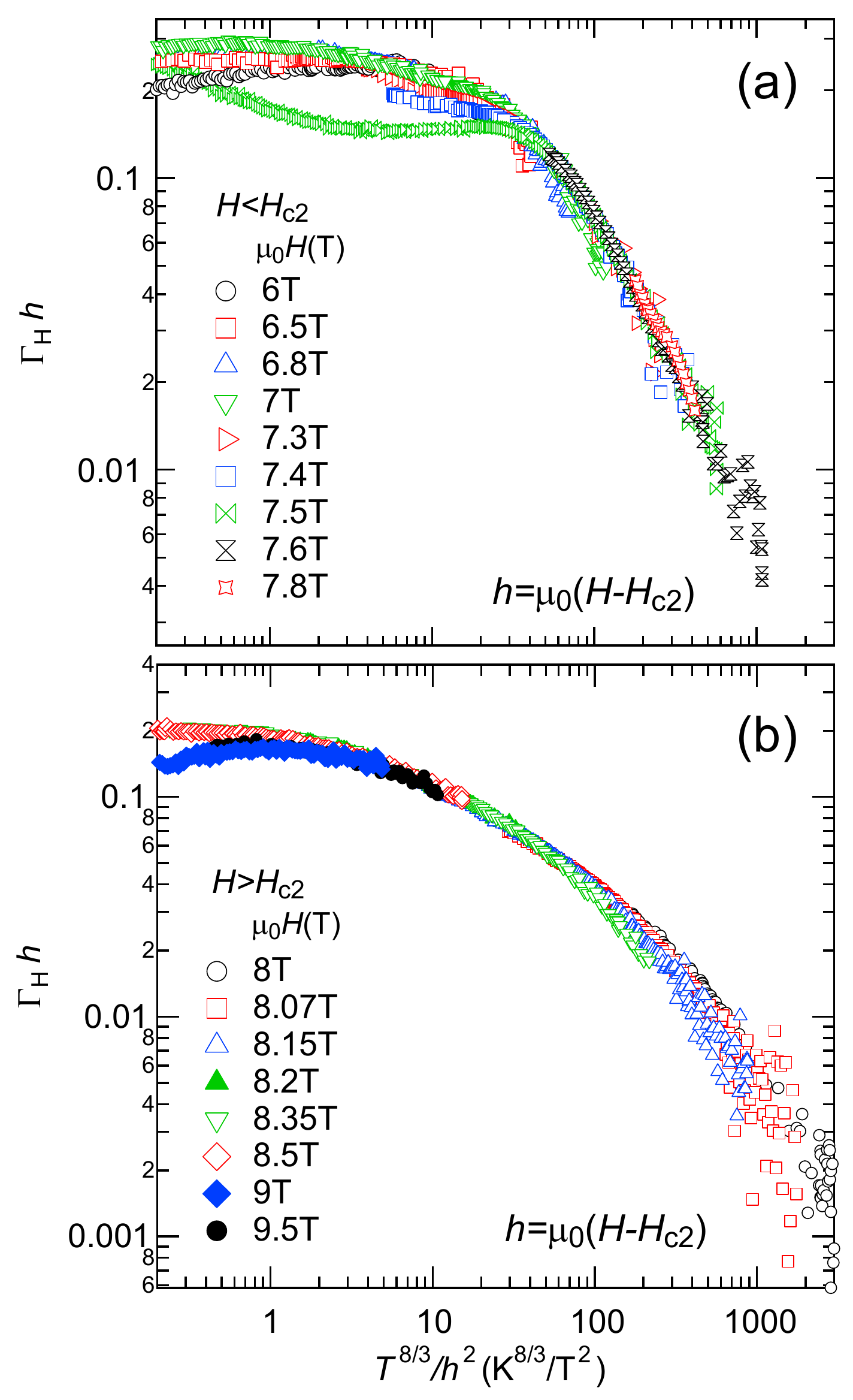}
\caption{Same plot as Fig.~5 in the main text, now also including all data points outside the QC2 scaling regime. The parameter $h=\mu_0(H-H_{c2})$ with $H_{c2}=7.845$~T.}
\label{scaling_all}
\end{figure}

The phase diagram displayed in Fig. 1, main text, illustrates the range in $T$-$H$ parameter space where quantum critical scaling in the magnetic Gr\"uneisen parameter with respect to the first (QC1) and second (QC2) quantum critical end point holds, respectively. It also shows a regime "QC1+QC2" where scaling fails because of the superposition of criticality from both instabilities. Furthermore scaling fails near the spin-density-wave states A and B and the yet unidentified regime C. In this section we focus on the scaling and its failure of $\Gamma_H$ due to QCEP1, while in the subsequent section respective analysis for QCEP2 is detailed.

As illustrated in Figs. 2 and 3 (inset) of the main text, at 0.2~K, specific heat and $\Gamma_H$ follow the field dependences in accordance with the scaling predictions for QCEP1. Here we analyze up to which temperature this scaling holds. For this purpose, Figure~\ref{1_G_H} shows a plot of 1/$\Gamma_H$ vs $\mu_0H$ at several elevated temperatures. Since for field tuned instabilities $\Gamma_H\sim (H_c-H)^{-1}$ with universal (temperature independent) pre-factor, such plot is best suited to determine the upper bound in temperature of the scaling regime. It is found, that all curves up to 0.6~K follow a universal dependence indicated by the blue line, while the data at 1~K follow a clearly different slope (incompatible with the expected universality). In addition they would extrapolate to a very different critical field, but the critical field should be temperature independent. Thus, scaling with respect to QCEP1 breaks down in between 0.6 and 1~K.

Next, we show scaling plots, similar to those of Fig. 5 in the main text, but now for the critical field of QCEP1, $H_{c1}=7.53(2)$~T. Figure.~\ref{scalingHc1} (a) shows the scaling analysis using $H_{c1}$ for fields below the critical field, while part (b) displays data for $H>H_{c1}$. The latter case is restricted to a very narrow field interval, because in the approach of $H_{c2}=7.845(5)$~T the magnetic Gr\"uneisen parameter changes sign due to the influence of QCEP2. Apparently, the data do not collapse at all for $H>H_{c1}$, which is attributed to the influence of QCEP2. In addition for $H<H_{c1}$ a data collapse is only found at low temperatures, consistent with the above observation of a breakdown of scaling at $T>0.6$~K. 

\section{Scaling analysis for QCEP2}

We now turn to the determination of the range in $T$-$H$ parameter space where scaling with respect to QCEP2 works (cf. regime "QC2" in Fig. 1 main text). For this purpose, we re-plot Fig. 5 of the main text, but now include all data to discuss where scaling fails.

In Figure~\ref{scaling_all} (a) deviation from scaling arises due to the presence of QCEP1 at fields below 7.5~T for temperatures below about 1~K. The very good data collapse at large temperatures indicates, that QCEP1 is clearly dominant at elevated temperatures for all fields $H<H_{c2}$.

Figure~\ref{scaling_all} (b) indicates that for fields between 9 and 9.5~T deviation from scaling is found at low temperatures, which could be associated with the regime "C".

To summarize, the magnetic Gr\"uneisen ratio indicates multiple quantum criticality with a QCEP1 near $\mu_0H_{c1}=7.53(2)$~T and a QCEP2 near $\mu_0H_{c2}=7.845(5)$~T. Using scaling analysis with respect to both $H_{c1}$ and $H_{c2}$, we determined the regimes "QC1" and "QC2" in the phase diagram of Fig. 1, main text, as well as the regime labeled "QC1+QC2" in which latter scaling fails because of the superposition of critical behavior from both instabilities.

\end{document}